\documentclass[twocolumn,english,aps,amsmath,amssymb,pra,superscriptaddress]{revtex4-1}

\usepackage[T1]{fontenc}
\usepackage[latin9]{inputenc}
\usepackage{lmodern}
\usepackage{array}
\usepackage{multirow}
\usepackage{amsbsy}
\usepackage{amstext}
\usepackage{amssymb}
\usepackage{graphicx}
\usepackage{esint}

\makeatletter

\providecommand{\tabularnewline}{\\}

\usepackage{braket}
\newcommand{\1}{{\rm 1\hspace{-0.9mm}l}}

\makeatother

\usepackage{babel}
\begin{document}
\title{Contrast in Multipath Interference and Quantum Coherence}

\author{Kai von Prillwitz}
\affiliation{Freiburg Institute for Advanced Studies, Albert-Ludwigs University
of Freiburg, Albertstra\ss e 19, 79104 Freiburg, Germany}

\author{\L{}ukasz Rudnicki}
\affiliation{Freiburg Institute for Advanced Studies, Albert-Ludwigs University
of Freiburg, Albertstra\ss e 19, 79104 Freiburg, Germany}

\author{Florian Mintert}
\affiliation{Freiburg Institute for Advanced Studies, Albert-Ludwigs University
of Freiburg, Albertstra\ss e 19, 79104 Freiburg, Germany}
\affiliation{Department of Physics, Imperial College London, London SW7 2AZ, United Kingdom}

\begin{abstract}
We develop a rigorous connection between statistical properties of an interference pattern and the coherence properties of the underlying quantum state.
With explicit examples, we demonstrate that 
even for inaccurate reconstructions of interference patterns properly defined statistical moments permit a reliable characterization of quantum coherence.
\end{abstract}
\maketitle
\section{Introduction}

Interference resulting from quantum coherence causes an abundance of effects that contradict our classical intuition.
Most people would probably be inclined to negate both the interference of independent photons \cite{HOM} or of mesoscopic molecules \cite{Arndt1, Facchi}  if there was no clear experimental evidence for the existence of both effects;
and despite the fact that interference phenomena in quantum mechanical systems have been observed for more than a century,
we can still find large missing pieces in our understanding of quantum coherence.

The fact that a coherent superposition of at least two path alternatives (two state-vectors in a more general, abstract description) is necessary for an \textit{interference pattern} to emerge, and that the achievable contrast increases with the number of states that are coherently over-imposed is one of the best established notions of elementary physics.
However, going beyond this qualitative observation, our intuition is typically not able to  answer the question of how  many path-alternatives are needed to generate a particular interference pattern with reduced contrast.
The overall aim of this paper is to explore the information content stored in the interference pattern, and to develop a framework which
addresses the above question.

The formal definition of quantum coherence requires a set of mutually orthogonal states $\ket{j}$, $j=1,\ldots,d$ with respect to which coherence is defined.
In an interferometric situation these states would correspond to different path alternatives and the number of paths that are being taken coherently is often referred to in terms of the lateral coherence length.
In the case of molecular networks one is typically interested in the number of chromophores over which an excitation is coherently distributed, so that coherent delocalization is defined in terms of the excited states of the individual chromophores \cite{Greg, efficient};
and in transport theory or quantum thermodynamics this reference basis is given by energy eigenstates \cite{scully2003extracting,Terry}.

In general, a pure state $\ket{\Psi}$ is considered $k$-coherent in terms of a given set of basis states $\ket{j}$,
if at least $k$ of the amplitudes $\langle\Psi|j\rangle$ are non-vanishing.
Since decoherence processes which are unavoidably present in actual physical situations result in the deterioration of quantum coherence, the description in terms of mixed states or density matrices becomes necessary. As every mixed state $\rho$ can intuitively be understood as an average over pure states,
 averaging over  \textit{incoherent} states $\ket{\Upsilon_i}$ will not result in any interference phenomena.
Consequently, any mixed state $\rho=\sum_ip_i\ket{\Upsilon_i}\bra{\Upsilon_i}$ that can be decomposed into a mixture of incoherent states with $p_i\ge 0$ is considered incoherent.
Analogously, every mixed state that can be expressed in terms of an average over pure states with no more than $k$-coherence is {\em not} $k+1$-coherent.
This motivates the commonly employed definition  \cite{FF,CatGS,Girolami} that a mixed state $\rho^{(k)}$ is $k$-coherent, if any ensemble $\{\ket{\Psi_i}\}$ that satisfies $\rho^{(k)}=\sum_ip_i\ket{\Psi_i}\bra{\Psi_i}$ for some set of probabilities $p_i$ contains at least one $k$-coherent state vector. The notion of $k$-coherence is thus similar to the concept of multipartite entanglement, since a mixed state is called $k$-partite entangled if any of its ensemble decompositions involves at least one $k$-partite entangled pure state \cite{Hor}.

Quantum coherence
has recently been recognized as a resource \cite{Biology,Plenio,Girolami,Aberg,Adesso,Terry} in the sense that
there are processes whose realization is facilitated by the consumption of coherence.
Various tools known from  the entanglement theory have thus been adapted for the classification and quantification of quantum coherence \cite{Fede,Plenio,FF,2015arXiv150205876S}.
Reconstruction of the complete density matrix is required to assess most of these tools,
and only few schemes work with fewer observables to be measured \cite{Girolami,geometric}.
On the one hand, this fact poses a rather high threshold for the analysis of coherence in laboratory experiments,
and, on the other hand, the abstract nature of the aforementioned tools limits the intuition that might be gained from their use.

\begin{figure}[h]
\includegraphics[width=1\columnwidth]{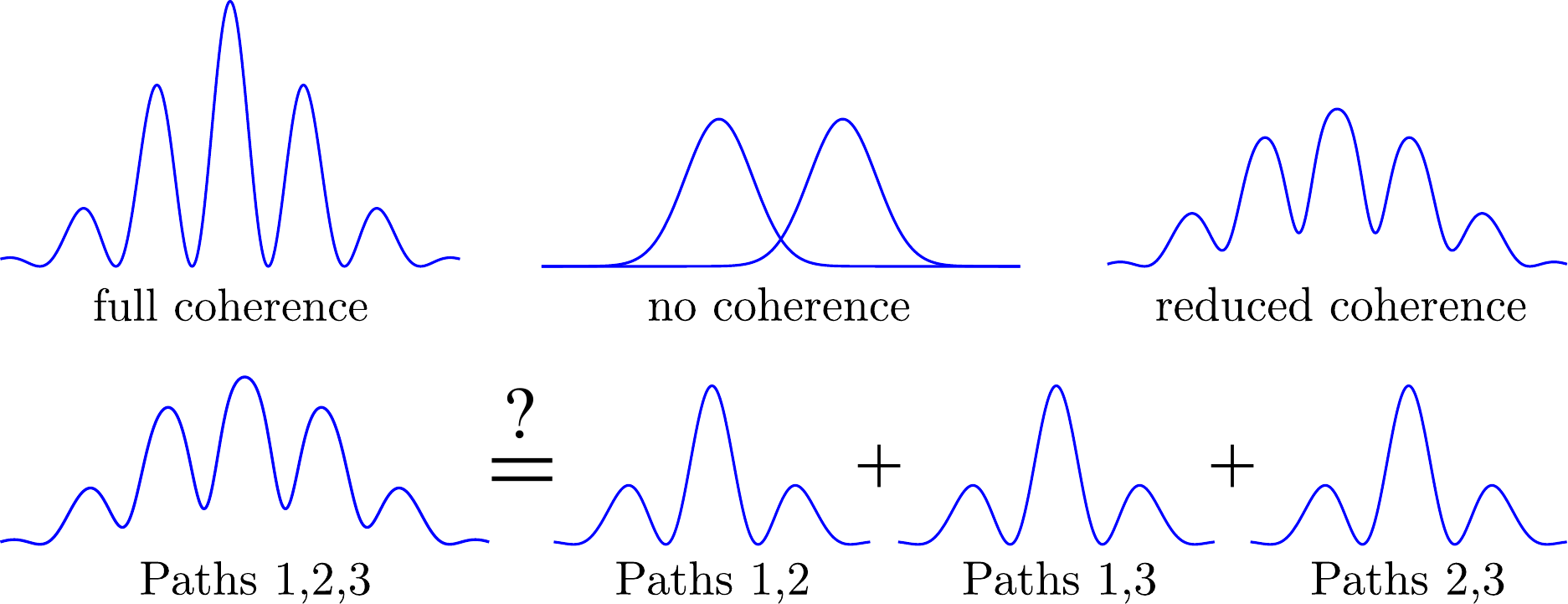}
\caption{Upper part: interference patterns corresponding to different types of coherence in  mixed-states. Lower part: an interference pattern that results from a state distributed over three path alternatives. Reduced coherence between the three paths can sometimes be expressed as the sum of two-path interference patterns. In that case the pattern can result from a two-coherent state.
\label{fig:interferencemixed}}
\end{figure}

We will strive for the identification of quantum coherence based on the interference pattern only.
If an interference pattern can be decomposed into a sum of patterns resulting from $k$-path interference,
then, this pattern does not permit to conclude on $k+1$-coherence (as exemplarily depicted in Fig.~\ref{fig:interferencemixed} for $d=3$ and $k=2$).
We will therefore identify (in Section II) properties of interference pattern asserting that such a decomposition is not possible.
In Section III we provide numerical evidence underlining the performance of the tools developed in Section II. The numerical studies have been designed to capture major practical issues, such as difficulties with the proper identification of the maximum of a complicated interference pattern, or its coarse graining.

\section{The interference pattern as a coherence classifier}\label{sec2}

\begin{figure}[h]
\centering{}\includegraphics[width=1\columnwidth]{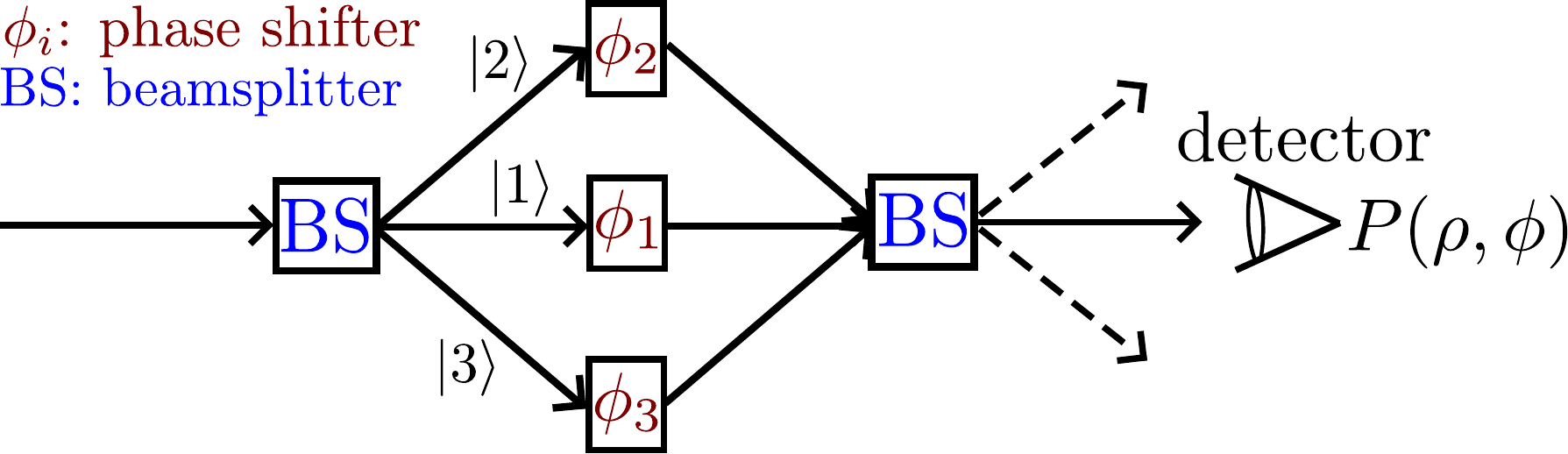}
\caption{Three--path extension of the Mach--Zehnder interference setup. An incoming particle is splitted into three contributions. After passing the individual phase shifters the three beams
are recombined at the second beam splitter where they interfere.\label{fig:interference}}
\end{figure}

We consider a rather general physical situation in which a superposition of different states is being established, and a certain level of decoherence results in the fact that this superposition is not perfectly coherent.
A specific realization of such a situation would be a Mach-Zehnder type of interferometer, as schematically depicted in Fig. \ref{fig:interference} for $d=3$, where the different path-alternatives define the basis states $\Ket{j}$.
An incoming object impinges on a beam-splitter (BS), that creates a coherent superposition of the basis states.
The phase shifters $\phi_i$ 
permit to generate an interference pattern that can be read off, once the object has crossed the second beam splitter.

The interference pattern 
\begin{equation}
P\left(\rho,\boldsymbol{\phi}\right)=\Braket{\Phi|\rho|\Phi}\equiv1+\sum_{j\neq m}\rho_{jm}e^{i\left(\phi_{j}-\phi_{m}\right)},\label{eq:definition_of_interference_pattern}
\end{equation}
with $\Ket{\Phi}=\sum_{j=1}^{d}e^{-i\phi_{j}}\Ket{j}$ is defined as the normalized probability distribution to observe an object in the output mode,
where $\rho$ is the state before crossing the second beam splitter.
Since on average only one out of $d$ objects exits through the output mode,
the interference pattern is given in terms of an over-normalized state with $\langle\Phi|\Phi\rangle=d$.

The simplest case of $d=2$ path alternatives corresponds to the original Mach-Zehnder-Interferometer, in which one can record the interference pattern  by tuning a single phase shifter. In general, the interference pattern is obtained by tuning $d-1$ phase shifters.
Beyond the obvious increase of dimensionality, also the structure of the pattern typically gets more complicated with growing $d$ since the dependence of the detection probabilities on the phases $\phi_i$ gets more sensitive. Our aim shall thus be to capture a more global part of the desired information, which is robust against small deviations of the tuned parameters. To this end we examine  various \textit{moments} of the interference pattern in question.

\subsection{Moments of the interference pattern \label{sub:Generalized-moments}}

One can certainly obtain some information on coherence from
the maximum of the interference pattern 
\begin{equation}
\max_{\boldsymbol{\phi}}P\left(\rho,\boldsymbol{\phi}\right)\ ,
\label{max}
\end{equation}
as its value, when larger than $k-1$, unambiguously identifies $\rho$ to be $k$-coherent.
In practice, however, this is not necessarily the best choice.
In particular, for highly coherent states, the interference pattern is a rapidly oscillating function so that optimizations will often  identify only local maxima with a resulting under-estimation of coherence properties.
Since, again, for highly coherent states, the optimum is given by a very narrow peak, an extremely accurate reconstruction of the interference pattern becomes necessary. 

A much more practical alternative would be to employ the uniform statistical moments
\begin{equation}
m_{n}=\int_{0}^{2\pi}\frac{d^{d}\boldsymbol{\phi}}{\left(2\pi\right)^{d}}P^n\left(\rho,\boldsymbol{\phi}\right)\equiv\int_{0}^{2\pi}\frac{d^{d}\boldsymbol{\phi}}{\left(2\pi\right)^{d}}\Braket{\Phi|\rho|\Phi}^{n}\ .\label{uniform}
\end{equation}
The first moment $m_{1}=1$ is just the norm of the interference pattern,
but the higher moments carry non-trivial information.
One would expect that increasing the order of moments improves the identification of $k$-coherence, because taking the limit $\lim_{n\rightarrow\infty}\left(m_{n}\right)^{1/n}$ is equivalent to finding the maximum (\ref{max}). On the other hand the required accuracy of the reconstructed pattern ({\it e.g.} from experimental data) necessary to assess a moment grows with $n$ \cite{ImageMoments}.

We thus strive for an approach that is based on moments of reasonably low order, which are more robust against small deviations of the interference pattern.
In order to find a good compromise between a sensitive identification and robustness with respect to imperfections, we utilize
the  \emph{generalized moments}
\begin{equation}
Q_{n}=\int d^{d}\boldsymbol{\phi}\, F\left(\boldsymbol{\phi}\right)P^n\left(\rho,\boldsymbol{\phi}\right)\ ,
\label{eq:definition_of_p_n}
\end{equation}
defined in terms of a suitably chosen $d$-dimensional probability distribution
$F\left(\boldsymbol{\phi}\right)$.
The simplest case $F\left(\boldsymbol{\phi}\right)=(2\pi)^{-d}$ reproduces the uniform moments Eq. (\ref{uniform}).
In the opposite case, when $F\left(\boldsymbol{\phi}\right)$ is strongly localized around the maximum of the interference pattern,
the value of the generalized moment $Q_n$ approximates the maximum $\max_{\boldsymbol{\phi}}P^n$.
This specific choice calls for the search of an optimal probability distribution which might be flawed by the same issues as encountered for Eq.~\eqref{max}.
With a sufficiently wide distribution $F$, on the other hand, the optimization landscape is substantially flatter than in Eq.~\eqref{max} what eases the optimization a lot. Using the generalized moments of low order together with the distribution $F$ encoding additional information, namely the expected position of the maximum of the interference pattern, one can reasonably merge advantages of both interference peaks (Eq.~\eqref{max}) and regular statistical moments (Eq.~\eqref{uniform}), avoiding complications brought by each of two approaches alone.

\subsection{The wrapped normal distribution}

Since the interference pattern defined in Eq.~\eqref{eq:definition_of_interference_pattern} is given in terms of all the phases $\phi_j$ as independent variables,
it is reasonable to define $F\left(\boldsymbol{\phi}\right)$ in terms of independent distributions for each phase,
{\it i.e.} $F\left(\boldsymbol{\phi}\right)=\prod_{j=1}^{d}f\left(\phi_{j};\mu_j\right)$.
The variable $\phi_{j}$ is thus distributed according to
$f\left(\phi_{j};\mu_j\right)$, 
with $f\left(\cdot\right)$ having
the same functional form for all $\phi_j$. By $\mu_j$ we denote the
expectation value of each of these distributions.
Although strictly speaking not necessary, we will assume equal width of all distributions and denote their standard deviation by $\sigma$.

An evaluation of Eq.~\eqref{eq:definition_of_p_n} requires the construction of so-called \emph{trigonometric moments}
$\Theta_{n}(\mu)\equiv\int_{0}^{2\pi}d\phi\, f\left(\phi;\mu\right)e^{in\phi}$ \cite{jammalamadaka} defined for any integer $n$.
Due to the fact, that the phases $\phi_j$ are defined only in an interval of width of $2\pi$, this step can be done explicitly for most typically employed distributions like the Lorentz or Gauss distributions.
To this end, it is helpful to take advantage of a {\em wrapped} version of a distribution \cite{jammalamadaka}.
In the case of the wrapped normal distribution,
the trigonometric moments are equal to the characteristic function
of the normal (unwrapped) distribution evaluated at integer arguments,
 $\Theta_{n}\left(\mu\right)=e^{in\mu}R_{n}$,
where $R_{n}=e^{-n^{2}\sigma^{2}/2}$.
With the help of the function $\Theta_{n}(\mu)$
one can perform the integration in Eq.~\eqref{eq:definition_of_p_n} and express the generalized moments as
\begin{eqnarray}
Q_{n}
&=&\!\!\!\!\!\sum_{i_{1},i_{2},...,i_{2n}=1}^{d}\!\!\!\!\!\rho_{i_{1}i_{n+1}}\rho_{i_{2}i_{n+2}}...\rho_{i_{n}i_{2n}}\prod_{j=1}^{d}\Theta_{n_{j}}\left(\mu_{j}\right),\label{eq:final-expression-for-p_n}\\
&&\mbox{with}\hspace{.5cm}n_{j}=\sum_{l=1}^{n}\delta_{j,i_{l}}-\sum_{l=n+1}^{2n}\delta_{j,i_{l}}\ .
\label{eq:n_i}
\end{eqnarray}

As argued above, the use of low order moments is desirable. We will therefore focus in the following on $n=1$, $n=2$ and $n=3$.
There is however no fundamental obstacle for generalizations to higher values of $n$.

\subsection{Threshold values}

Before one can use the moments defined in Eqs.~\eqref{eq:definition_of_p_n} or \eqref{eq:final-expression-for-p_n} to rigorously identify coherence properties,
one needs to find the maximum that $Q_n$ can adopt for $k$-coherent states.
In the present case such an optimization can be done explicitly, confirming that the maximum among all $k$-coherent states is provided by
\begin{equation}
\ket{W_k}=\frac{1}{\sqrt{k}}\sum_{j=1}^k e^{-i\varphi_j}\ket{j}\ ,\label{maxst}
\end{equation}
{\it i.e.} a perfectly balanced coherent superposition of $k$ basis states.

To arrive at this conclusion, one may first realize that the generalized moments $Q_n$ are convex, {\it i.e.}
\begin{equation}
Q_n(\eta\rho_1+(1-\eta)\rho_2)\le \eta Q_n(\rho_1)+(1-\eta)Q_n(\rho_2)\ ,
\label{eq:convex}
\end{equation}
for $0\le\eta\le 1$ and any pair of
density matrices $\rho_1$ and $\rho_2$.
This is a direct consequence of the two facts that the $n_{\textrm{th}}$ power of a linear functional like $\Braket{\Phi|\rho|\Phi}^{n}$ is convex for $n\ge 1$, and that the integral 
$\int d^{d}\boldsymbol{\phi}  F\left(\boldsymbol{\phi}\right)$
preserves convexity as $F$ is non-negative.
Since states $\rho^{(k)}$ that are at most $k$-coherent (for any value of $k$) define a convex set
({\it i.e.} $\eta\rho_1^{(k)}+(1-\eta)\rho_2^{(k)}$ is no more than $k$-coherent)
the maximum of $Q_n$ over $k$-coherent density matrices is always reached for a pure state.

The most general $k$-coherent pure state reads
\begin{equation}
\ket{\Psi^{(k)}}=\sum_{j=1}^k\sqrt{\lambda_j}e^{-i\varphi_j}\ket{j},\mbox{with }\lambda_j\ge 0\ , \label{kcohpure}
\end{equation}
assuming (without loss of generality) that exactly the first $k$ basis states are comprised with non-vanishing weights $\lambda_j$ in the coherent superposition.
In Appendix \ref{sec:Derivation-of-threshold-values} it is shown that the optimization of the phase factors $e^{-i\varphi_j}$ can be performed independently of the optimization over the real amplitudes $\lambda_j$,
and that the maximum is obtained if the $\varphi_j$ coincide with the expectation values $\mu_j$ of $f(\phi_j)$.
Also the remaining optimization over the $\lambda_j$ can be performed very generally.
As further shown in Appendix \ref{sec:Derivation-of-threshold-values}, the quantity to be optimized is a Schur-concave function which is maximized for $\lambda_j=1/k$ for $j=1,\hdots,k$.

Summarizing the above considerations, the maximum of $Q_n$ that can be adopted for $k$-coherent states with a given distribution $F$ reads
\begin{equation}
Q_{n}^{\left(k\right)}=\max_{\rho^{(k)}} Q_n(\rho^{(k)})=k^{1-n}\sum_{l=0}^{n^{2}}v_{l}^{\left(n,k\right)}e^{-l\sigma^{2}},\label{maxbound}
\end{equation}
with the coefficients $v_{l}^{\left(n,k\right)}$ given in Table \ref{Table1}.
Any excess of this threshold value is an unambiguous identification of coherence properties beyond $k$-coherence.

\begin{table}[h]
\begin{tabular}{p{1cm} p{1.5cm} p{1.7cm} p{3.8cm}}
\hline \hline
$l$ \quad& $n=1$ & $n=2$ & $n=3$\tabularnewline
\hline 

0 & $1$ & $2k-1$ & $4-9k+6k^{2}$\tabularnewline
%\hline 
1 & $K_{1}$ & $4$ & $3K_{1}[11+3k(2k-5)]$\tabularnewline
%\hline 
2 & - & $K_{1}K_{2}K_{3}$ & $9K_{1}K_{2}^{2}K_{3}$\tabularnewline
%\hline 
3 & - & $2K_{1}K_{2}$ & $K_{1}K_{2}^{2}(45+kK_{10})$\tabularnewline
%\hline 
4 & - & $K_{1}$ & $3K_{1}[k(55+2kK_{9})-52]$\tabularnewline
%\hline 
5 & - & - & $9K_{1}K_{2}K_{3}$\tabularnewline
%\hline 
6 & - & - & $2K_{1}K_{2}K_{3}$\tabularnewline
%\hline 
7 & - & - & $6K_{1}K_{2}$\tabularnewline
%\hline 
8 & - & - & $0$\tabularnewline
%\hline 
9 & - & - & $K_{1}$\tabularnewline
\hline \hline
\end{tabular}\caption{The coefficients  $v_{l}^{\left(n,k\right)}$ that characterize the threshold values in Eq.~\eqref{maxbound} for $n=1$, $n=2$ and $n=3$. The index $l$ runs from $0$ to $n^2$ and $K_{m}=k-m$ is a short hand notation.}\label{Table1}
\end{table}

In the case of the lowest moment $Q_1$, one obtains
\begin{equation}
Q_1(\rho)=1+e^{-\sigma^2}\sum_{i_{1}\neq i_{2}}\rho_{i_{1}i_{2}}\prod_{j=1}^{d}e^{in_{j}\mu_{j}}\ ,\label{first moment}
\end{equation}
with $n_{j}\in\left\{ -1,0,1\right\}$.
That is, the width $\sigma$ of the distribution enters only as multiplicative factor $e^{-\sigma^2}$.
Since the threshold values for $k$-coherence and the values of $Q_1$ for any quantum state scale with $\sigma$ in exactly the same fashion,
the question of whether $Q_1(\rho)$ exceeds a threshold value is independent of the width of the utilized distribution $F$.
For $n=1$ there is thus no advantage of generalized moments over uniform moments.
As we will see in the following, however, the former are clearly advantageous for higher moments $Q_n$ with $n>1$.

\section{Numerical analysis\label{sec:Comparison-of-different-moments}}

Having established the rigorous properties of the generalized moments
$Q_{n}$, it remains to identify the range of optimal values of $\sigma$.
While narrow distributions require an accurate identification of the
maximum in the interference pattern, they yield strong criteria if
this maximum is found. On the other hand wider distributions give
robust but potentially weaker criteria. 

In the numerical studies, we thus address the following three questions:
\begin{itemize}
\item[-]how many states are detected as $k$-coherent by the $Q_n$?
\item[-]how sensitive are the $Q_n$ with respect to mis-estimating the maxima of the interference pattern? 
\item[-]how does the sensitivity of the $Q_n$ depend on the width $\sigma$ of the distribution $F$?
\end{itemize}

To address these questions, we will consider ensembles of density matrices and determine how many states are detected to be \emph{k}-coherent.
We will characterize the performance of the $Q_n$ in terms of the
\emph{detection ratio} $R$ defined as the ratio number of states detected as \emph{k}-coherent and the ensemble size.
Since there is no reliable construction of \emph{k}-coherent states any ensemble will always contain states that are {\em not} \emph{k}-coherent; the maximally achievable detection ration can therefore be substantially smaller than $1$, so that its absolute value is {\em not} a good indicator of the strength of $Q_n$.
To obtain an estimate of the overall strength, we will thus first compare the $Q_n$ with previously known tools for the characterization of \emph{k}-coherence and subsequently investigate the dependence of the detection ratio on errors in estimating the maximum of the interference pattern and the width $\sigma$.

\subsection{Comparison with other criteria}

In order to estimate the strength of the generalized moments $Q_n$
we employ a comparison with a
hierarchy of separability criteria that detect \emph{k}-coherence \cite{Fede,Bjorn}.
We shall perform this comparison based on an ensemble of $1000$ states that are stationary solutions of a driven, dissipative disordered network \cite{Bjorn}.

\begin{table}[h]

\begin{centering}
\begin{tabular}{ p{1.9cm}p{1.5cm}p{1.5cm} p{1.5cm} p{1.5cm}}
\hline\hline
& $k=2$ & $k=3$ & $k=4$ & $k=5$\tabularnewline
\hline 
hierarchy & 1000 & 989 & 562 & 12\tabularnewline
moments & 1000 & 969 & 347 & 0\tabularnewline
\hline \hline
& \multicolumn{3}{c} {number of states for which:} & \tabularnewline
& $k_m>k_h$ & $k_m<k_h$ & $k_m=k_h$ & \tabularnewline 
\hline
& 83 & 318 & 599 & \tabularnewline 
\hline \hline
\end{tabular} 

\caption{
Comparison between the hierarchy of separability criteria and the generalized moments with an asymptotically peaked distribution ({\it i.e.} $\sigma\to 0$).
Upper part: Number of states detected as \emph{k}-coherent for $k=2,3,4,5$.
Lower part: Number of states for which the generalized moments detected strictly larger, smaller and the same $k$ as the hierarchy.
\label{tab:Comparison_to_hierarchy}}

\par\end{centering}

\end{table}

In Table \ref{tab:Comparison_to_hierarchy} (upper part) we list the number of
states that have been confirmed to be \emph{k}-coherent for $k=2,3,4,5$, with the help of the
hierarchy and the generalized moments $Q_n$ with $\sigma\to 0$ respectively.
With either method all states are identified as $2$-coherent;
since only diagonal states are incoherent, the identification of $2$-coherence is not challenging and it comes at no surprise that both methods perform well.
In general, the moments perform particularly well in detecting low $k$-coherence, while for $k=4,5$ the superiority of the hierarchy becomes more evident. 
A state-wise comparison (lower part of Table \ref{tab:Comparison_to_hierarchy}), however, reveals that there is also a significant number of states ($83$ out of $1000$) for which the moments detected larger $k$ than the hierarchy. Most of these states have been identified as $4$-coherent by the moments and $3$-coherent by the hierarchy. 
Thus, even though the hierarchy is overall slightly stronger than the generalized moments, the results suggest that the performance of the moments is at least comparable to those of other known tools.
This is striking since prior tools are designed to employ the information content of the full density matrix whereas the generalized moments are based on limited, easily accessible information.

\subsection{Dependence on distribution and errors}

Having verified the overall strength of the $Q_n$, we are in the position to investigate how this strength depends on the width $\sigma$ of the distribution $f$ and errors in placing the centers of the distributions.
To quantify the error in determination of $\boldsymbol{\phi}^{\mathrm{max}}$ providing
the maximum of the interference pattern, we
assume that the centers $\boldsymbol{\mu}$ of the distributions are
shifted away from the maximum by some vector $\boldsymbol{\delta}$.
In our simulations we can numerically find $\boldsymbol{\phi}^{\mathrm{max}}$
(in the first considered ensemble of states we have by construction
$\boldsymbol{\phi}^{\mathrm{max}}=0$), randomly draw $\boldsymbol{\delta}$
and in the last step set $\boldsymbol{\mu}=\boldsymbol{\phi}^{\mathrm{max}}+\boldsymbol{\delta}$.

 The vector $\boldsymbol{\delta}$ emulates the general effect of inaccurate determination of the maximum. The latter, however, does not necessarily need to be  caused by optimization issues;
since the maximum of a measured interference pattern can never be determined with a precision better than that allowed by the coarse graining (binning) of the measurement,
a finite $\boldsymbol{\delta}$ can also arise from experimental limitations.

If all the $\delta_{i}$ vanish, the distribution is centered around
the maximum of the interference pattern, and since the entire interference
pattern is invariant under a global phase shift, this holds also if
all the $\delta_{i}$ coincide. We therefore define the invariant
deviation vector $\tilde{\delta}_{i}=\delta_{i}-\bar{\delta}$, where
the term $\bar{\delta}=\sum_{i=1}^{d} \delta_{i}/d$ removes
the aforementioned ambiguity. All the $\tilde{\delta}_{i}$ are treated
as  normally-distributed independent random variables
\begin{equation}
\tilde{\delta}_{i}\sim\mathcal{N}\left(0,\sigma_{G}^{2}\right)\textrm{,}\label{eq:delta_i_tilde_distribution}
\end{equation}
characterized by a single width $\sigma_{G}$.
The particular choice $\sigma_{G}=0$ refers to the optimal case when
the maximum of the pattern is identified perfectly, while positive widths
cause random, but statistically controlled shifts.
In order to obtain the desired width in Eq.\eqref{eq:delta_i_tilde_distribution},
the primary, non-invariant parameters $\delta_{i}$ need to be distributed
according to $\mathcal{N}\left(0,\frac{d}{d-1}\sigma_{G}^{2}\right)$.
In other words, the global phase invariance provides the effective decrease
of the width by the factor $\sqrt{\left(d-1\right)/d}$, so that in
the limit of large $d$ the invariance in question does not result in observable effects.
%does not play any role. 

Since for a fixed choice of $\sigma_{G}$, the ability to identify
coherence is expected to depend on the specific choice of the $\delta_{i}$,
we average the detection ratio over a large number, say $N=1000$,
of deviation vectors $\boldsymbol{\delta}$,
and the averaged detection ratio is denoted further by $\left\langle R\right\rangle $.

\subsubsection{A single-parameter family of states}

\begin{figure}[h]

\begin{centering}
\includegraphics[width=1\columnwidth]{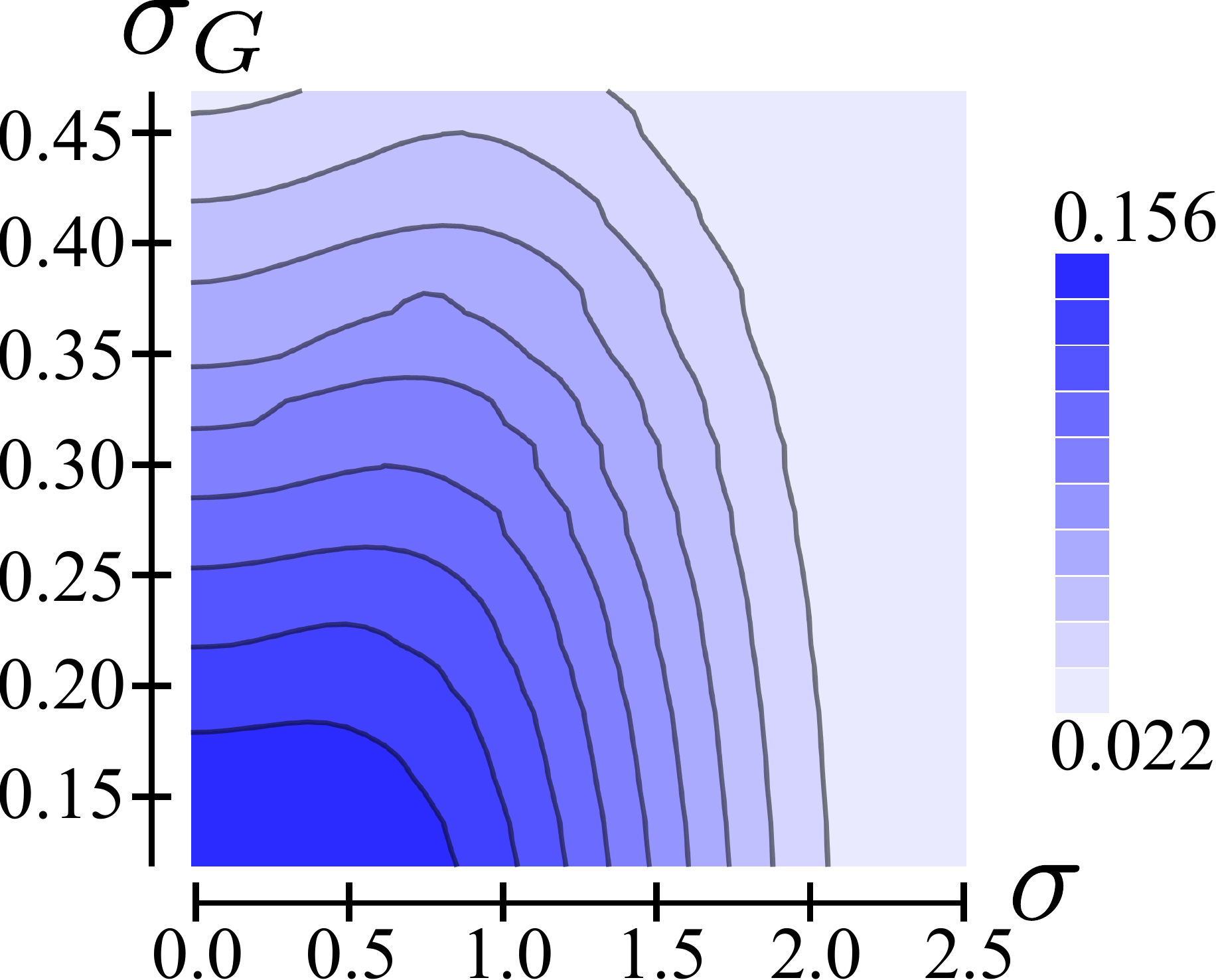}
\caption{Dependence of the average detection ratio $\left\langle R\right\rangle $ on width $\sigma$
and error $\sigma_G$ (indicated by the colorscale).
The depicted data refer to $Q_{3}$, $d=k=7$
and the ensemble given by Eq.~\eqref{eq:first_ensemble}.
There is a broad region ($\sigma_G\lesssim 0.15$ and $\sigma\lesssim 0.7$) in which $Q_3$ performs nearly as well as in the ideal case $\sigma_G= 0$ and $\sigma\to 0$. \label{fig:detection_ratio_plotsA}}

\par\end{centering}

\end{figure}

Let us start with a simple ensemble 
\begin{equation}
\rho_{a}=a\Ket{\Psi_{W}}\Bra{\Psi_{W}}+\frac{1-a}{d}\boldsymbol{1}\textrm{,}\label{eq:first_ensemble}
\end{equation}
with $\Ket{\Psi_{W}}=\sum_{i=1}^{d}\Ket{i}/\sqrt{d}$
and the identity matrix $\boldsymbol{1}$,
parametrized by the real parameter $0\leq a\leq1$. All $d$ basis states are populated
with equal weight independently of $a$; for $a=1$, $\rho_{a}$ describes
a perfectly coherent superposition, and $a=0$ corresponds to the
situation with no phase coherence. Here, we consider $d=7$ and
try to detect \emph{k}-coherence for $k=7$. For both parameters  $\sigma$
and $\sigma_G$ being fixed we numerically integrate over the range of the parameter $a$ in order to obtain the detection
ratio $R$. For any combination $\left(\sigma,\sigma_{G}\right)$
we further average $R$ over $1000$ deviation vectors $\boldsymbol{\delta}$. The averaged detection
ratio $\left\langle R\right\rangle $ for the third moment $Q_{3}$
as a function of $\sigma$ and $\sigma_{G}$ is shown in Fig. \ref{fig:detection_ratio_plotsA}.

\begin{figure}[h]

\begin{centering}
\includegraphics[width=1\columnwidth]{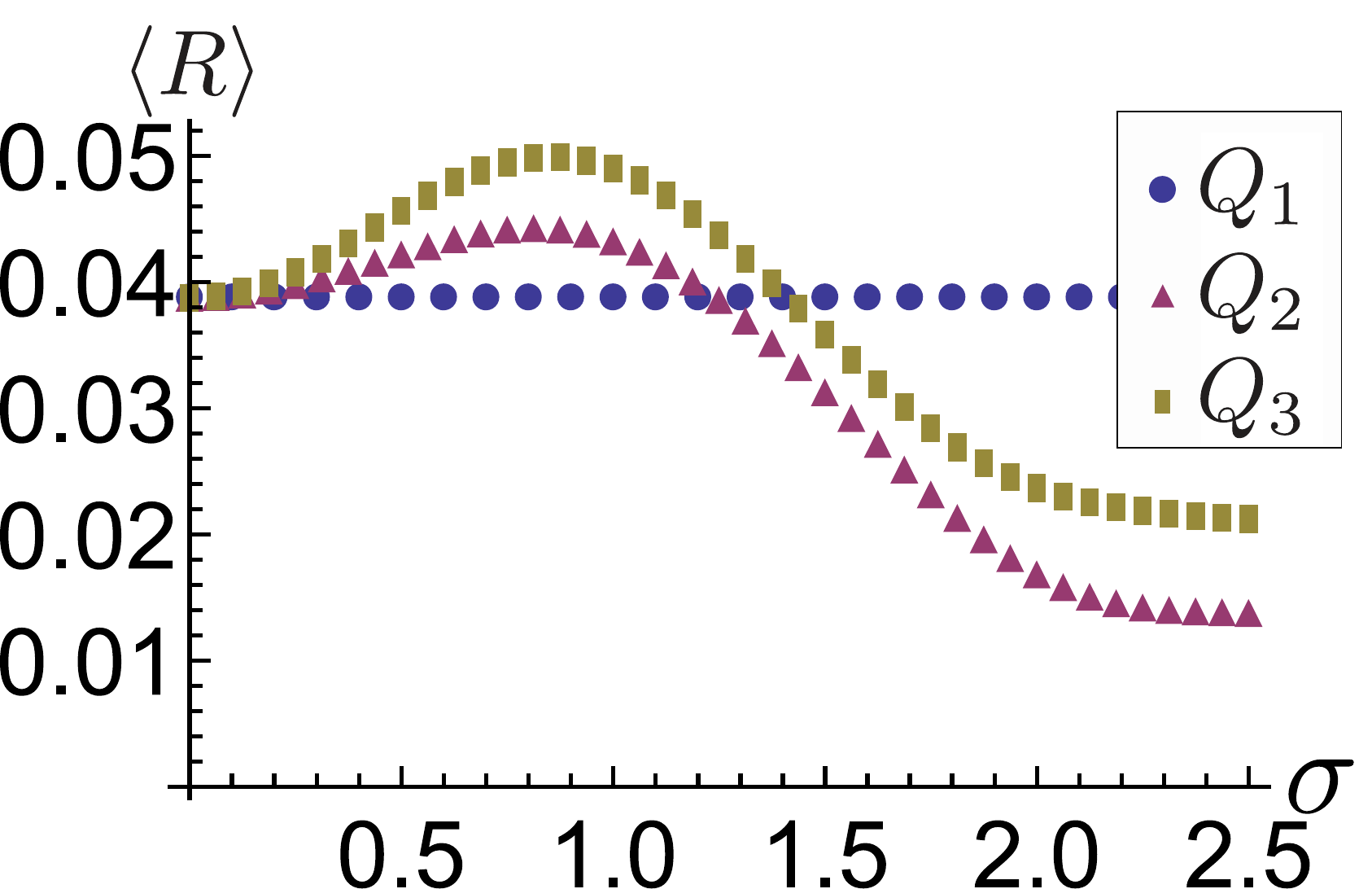}
\caption{Dependence of the average detection ratio $\left\langle R\right\rangle $ on  width $\sigma$ for $\sigma_{G}=0.4$.
As discussed above, the performance of $Q_1$ is independent of $\sigma$, but there is an optimal, finite $\sigma$ for $Q_{2}$ and $Q_{3}$ . \label{fig:detection_ratio_plotsB}}

\par\end{centering}

\end{figure}

As expected, when $\sigma$ is fixed, $\left\langle R\right\rangle $ decays monotonically
with $\sigma_{G}$. An interesting
feature can be captured while looking at the perpendicular direction, {\it i.e.} on the average detection ratio as a function of $\sigma$
with fixed $\sigma_{G}$. There is a broad range of $\sigma_{G}$,
for which $\left\langle R\right\rangle $ is
an increasing function of $\sigma$. In other words, broader distributions used during the construction of the generalized moments can compensate the inaccuracy in determination of the maximum.  Fig. \ref{fig:detection_ratio_plotsB} explicitly shows the above effect for an exemplary intersection
taken at $\sigma_{G}=0.4$. The largest value of $\left\langle R\right\rangle $
is in this case not achieved for $\sigma=0$, but for a finite width
$\sigma\approx0.9$.
In more detail, for $\sigma=0$ (exact evaluation of the supposed maximum of the interference pattern)
the average detection ratio for $\sigma_{G}=0.4$ is  $0.039$ while it equals  $0.166$ when $\sigma_{G}=0$ (supposed maximum is the true maximum). 
This observation indicates that
the moments with  narrower distributions are rather sensitive to the error in determination of
the maximum. In the suboptimal case the average detection ratio drops to $\approx23\%$
of the optimal value achieved for  $\sigma_{G}=0$. The larger width $\sigma=0.9$ (still for $\sigma_G=0.4$) provides an increase of $\left\langle R\right\rangle$ by a factor of $1.28$,
to the value of $0.050$.

\subsubsection{Random states}

In order to test the universal validity of the above observations
we performed a similar computation with arbitrary random states
$\rho=U\Lambda U^{\dagger}$, where $\Lambda$ is a diagonal matrix
describing the spectrum of $\rho$ while $U$ is a unitary transformation.
The matrix $U$ is drawn from the Circular Unitary Ensemble (CUE \cite{RUM, RUM2}), while $\Lambda$ contains squared absolute values of components
from a single column of a unitary matrix, which was also generated
with the help of CUE. The ensemble used to obtain the detection ratio contained
$500$ random states. Due
to substantial computational effort we averaged  the detection ratio (with fixed $\sigma$ and $\sigma_G$) over $25$ deviation vectors
$\boldsymbol{\delta}$.

 Let us denote by $\left\langle R\right\rangle _{\mathrm{ref}}$ the average detection ratio corresponding to the first moment. By construction, the same value of $\left\langle R\right\rangle$ is obtained while using any other moment with $\sigma=0$. 
As discussed in the context of Eq.~\eqref{first moment},
the performance of the first generalized moment is independent
of the parameter $\sigma$. We shall thus utilize  $\left\langle R\right\rangle _{\mathrm{ref}}$ as the reference quantity, and calculate 
\begin{equation}r_n=\left\langle R\right\rangle _{n}^{\mathrm{max}}/\left\langle R\right\rangle _{\mathrm{ref}},\end{equation} where by $\left\langle R\right\rangle _{n}^{\mathrm{max}}$ we denote the average detection ratio for the $n$th moment calculated with $\sigma=\sigma^{\mathrm{max}}$ being the value of $\sigma$, for which $\left\langle R\right\rangle$ based on the $n$th moment is maximal.
Finally, by $\left\langle R\right\rangle _{\mathrm{opt}}$ we denote the average detection ratio (independent of $n$)  obtained in
the optimal setting $\sigma_{G}=0=\sigma$. 

\begin{table}[h]

\begin{centering}
\begin{tabular}{p{0.8cm} p{1.2cm} p{0.8cm} p{1.2cm}p{1.2cm} p{1.2cm} p{1.2cm}}
\hline \hline
\emph{\vphantom{$\hat{R}$}k} & $\left\langle R\right\rangle _{\mathrm{opt}}$ & $\sigma_{G}$  & $\left\langle R\right\rangle _{\mathrm{ref}}$ &
$r_2$ &
$r_3$ & $\sigma^{\mathrm{max}}$\tabularnewline
\hline 
\multirow{2}{*}{$\geq4$} & \multirow{2}{*}{$0.55$} & $0.3$ & $0.45$ & $1.00$ & $1.01$ & $0.3$\tabularnewline
 &  & $0.8$ & $0.11$ & $1.10$ & $1.21$ & $0.7$\tabularnewline
\hline 
\multirow{2}{*}{$\geq5$} & \multirow{2}{*}{$0.19$} & $0.3$ & $0.10$ & $1.03$ & $1.05$ & $0.5$\tabularnewline
 &  & $0.8$ & $0.007$ & $1.21$ & $1.48$ & $0.8$\tabularnewline
\hline 
$\geq6$ & $0.01$ & $0.3$ & $0.002$ & $1.13$ & $1.21$ & $0.6$\tabularnewline
\hline\hline 
\end{tabular}\caption{Optimal (opt), reference (ref) and maximal ($r_n$, with respect to the reference value) average detection ratios for the first three generalized moments.
$\sigma^{\mathrm{max}}$ is the value of $\sigma$ for which $\left\langle R\right\rangle$ for both $n=2,3$ becomes maximal.
Several values of \emph{k} and $\sigma_{G}$ have been tested.  \label{tab:detection_ratios_random_states}}

\par\end{centering}

\end{table}

The results of our numerical analysis are presented in Table \ref{tab:detection_ratios_random_states}. Comparing $\left\langle R\right\rangle _{\mathrm{opt}}$ with the ratios
$\left\langle R\right\rangle _{\mathrm{ref}}$ in the suboptimal setting
(positive $\sigma_{G}$), we confirm the previous observation that the performance
of the moments (those with narrow distributions)
can be highly sensitive to uncertainty of $\boldsymbol{\phi}^\mathrm{max}$.
Similarly to the case of the first ensemble investigated, the second and third moment with positive
width $\sigma$ can significantly increase the detection ratio in
this suboptimal setting.

\section{Analysis for mixed states\label{sec:mixed_states}}
In Section \ref{sec2} we showed that the generalized moments $Q_n$ are convex with respect to the density matrix, what further implies that their maxima are provided by pure states (see Eq.~\eqref{maxst}).
The identification of the maximum value of $Q_n$ that can be taken for mixed states with given purity, thus allows to strengthen the detection of coherence in mixed states.
As we explicitly demonstrate here for the case $n=2$, there are however strongly mixed states that yield values of $Q_2$ close to the achievable maximum. That is, the present approach is by no means limited to pure or weakly mixed states, but, it can indeed identify $k$-coherence also for substantially mixed states.

In Appendix \ref{sec:Derivation-of-maximum}  we show that the global maximum 
\begin{equation}
\max_\rho (Q_2(\rho)\ |\ \textrm{Tr}\rho^{2}=\mathcal{P})
\label{eq:maxrhop}
\end{equation}
taken over all states with given purity $\mathcal{P}$ is obtained for the state
\begin{equation}
\varrho_\mathrm{max}\left(\mathcal{P}\right)=\left(1-\sqrt{\frac{\mathcal{P}d-1}{d-1}}\right)\frac{\1_{d}}{d}+\sqrt{\frac{\mathcal{P}d-1}{d-1}}\Ket{W_{d}}\Bra{W_{d}},\label{state1}
\end{equation}
with $\Ket{W_{d}}$ defined in Eq.~\eqref{maxst}.
Moreover, any state of the form Eq.~\eqref{state1} with purity 
\begin{equation}
\textrm{Tr}\rho^{2}\le\mathcal{P}_k=\frac{k^{2}-2k+d}{d\left(d-1\right)}\ ,\label{range1}
\end{equation}
is at most $k$-coherent.
$\mathcal{P}_k$ is thus the smallest purity that permits to identify $k+1$-coherence with the present moments.
%The value
%\begin{equation}
%Q_{2}(\varrho_\mathrm{max}(\mathcal{P}))
%\end{equation}
%thus provides the threshold value that indicates $k+1$-coherence for states with purity $\mathcal{P}\leq\mathcal{P}_{k}$.

\begin{figure}[h]
\includegraphics[width=1\columnwidth]{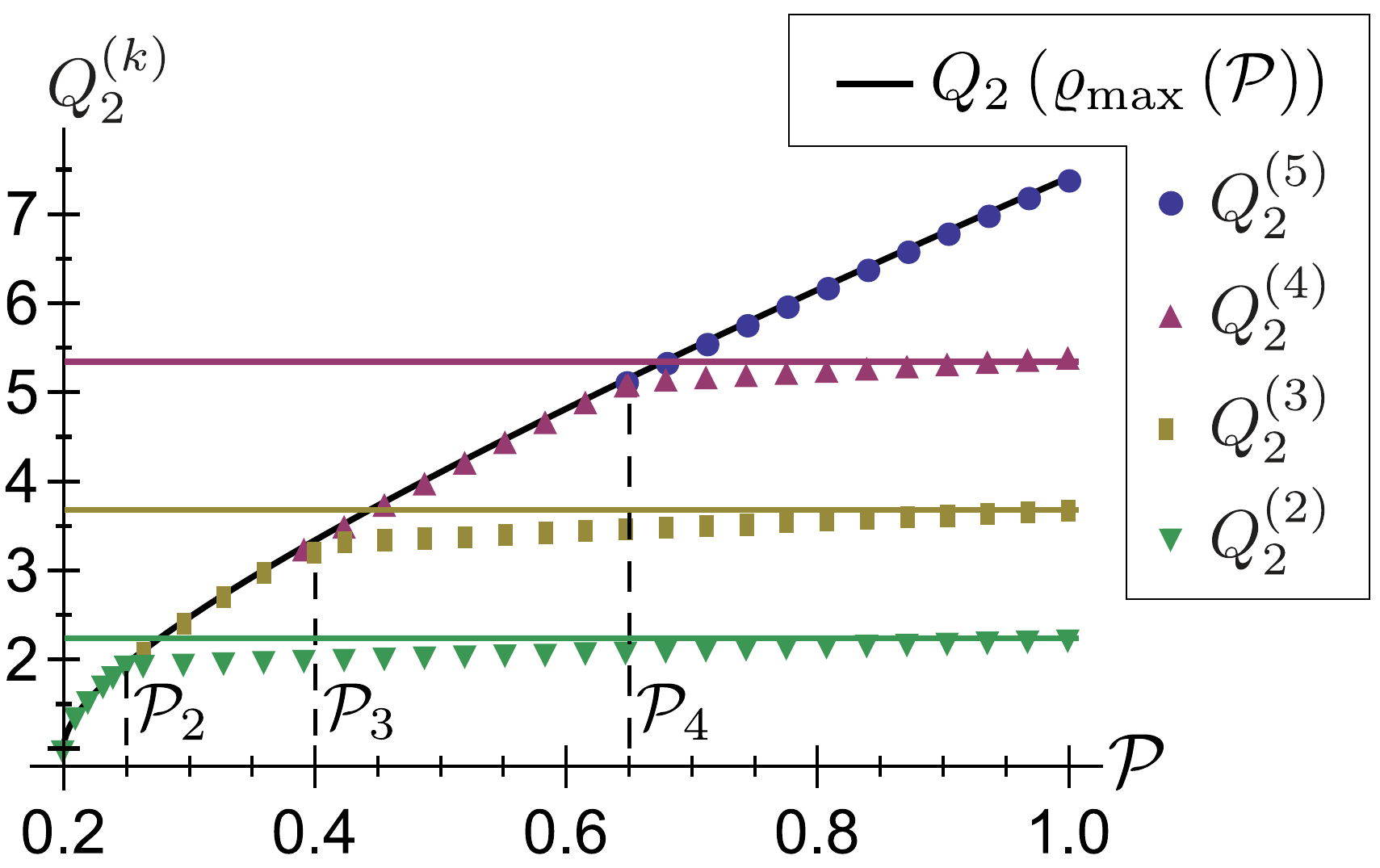}
\caption{The values $Q_{2}^{(k)}$
depict the numerically obtained, purity-dependent threshold values of $k$-coherence for $k=2,3,4,5$ in a $d=5$-dimensional system and $\sigma=1$. The solid curve denotes
the analytical result obtained for the state (\ref{state1}). The horizontal lines denote the original threshold values obtained by pure states. \label{fig:opt}}
\end{figure}

Fig.~\ref{fig:opt} depicts $Q_{2}(\varrho_\mathrm{max}(\mathcal{P}))$
as function of $\mathcal{P}$ (solid curve), and $\mathcal{P}_{k}$ for a $5$-dimensional system with $\sigma=1$.
The symbols (triangles, rectangles, circles) depict the numerically obtained maximum $Q_2^{(k)}$ of $Q_2$ with the maximization performed over all at most $k$-coherent states with given purity (see Eq.~\eqref{dec2} in the Appendix);
the horizontal lines depict the corresponding values for $\mathcal{P}=1$ to guide the eye.

As one can see, the values $Q_2^{(k)}$ coincide with $Q_{2}(\varrho_\mathrm{max}(\mathcal{P}))$ for $\mathcal{P}\leq\mathcal{P}_k$;
for $\mathcal{P}>\mathcal{P}_k$, $Q_2^{(k)}$ is nearly constant, {\it i.e.} there is a very small increase with $\mathcal{P}$. 
This means, that there are rather highly mixed $k+1$-coherent states that yield values above the threshold values of $Q_2^{(k)}$ for $k$-coherence.
The approach developed hitherto is thus able to identify coherence reliably even for rather strongly mixed states, even if no information on purity is available.
Since the range over which $Q_2^{(k)}$ is nearly constant, is the larger ({\it i.e.} including lower values of $\mathcal{P}$), the smaller $k$ is, this holds in particular for the identification of low $k$ coherence.
That is, in particular for $k\ll d$, (as it typically is the case in excitation transport \cite{rop_fede}),  
the present framework can detect coherence very well even for quantum states with substantial degree of mixing;
but if necessary, one may always resort to the purity-dependent threshold values in order to improve the detection.

\section{Conclusion}

As we have seen, an interference pattern permits to draw rigorous conclusions on coherence beyond the intuitive, qualitative expectation that an interference contrast grows with increasing coherence properties. 
From a practical point of view, the freedom in choice of sampling
as well as the possibility to include additional information (like purity) makes this approach flexible, so that it can be tailored for the specific properties of a system under investigation.
That is, limitations on experimentally variable quantities may be compensated through suitably chosen distributions with variable widths that reflect the realistically achievable sampling.
In particular, with noisy data that does not permit to reconstruct an entire interference pattern reliably, the generalized moments of low order still can characterize coherence properties reliably.

Here, we have been considering the case of $d$ independently adjustable phases, but the underlying framework can be generalized 
also for the variation of fewer phases.
Moreover, we developed the general framework
which can utilize a wrapped version of an arbitrary probability distribution. While
the normal distribution seems to be the most natural first choice, more sophisticated
distributions can even better support a particular experimental realization. Beyond the conceptual connection between a directly observable interference pattern and the underlying, abstract coherence properties, the present approach thus provides a versatile method to characterize coherence properties in a wide range of systems. 

At the end, let us establish a general link between the generalized moments discussed in this paper and the commonly employed characterization of quantum coherence in terms of the $l_{1}$-norm of coherence $C_{l_{1}}(\rho)=\sum_{i\neq j}|\rho_{ij}|$ \cite{Plenio}.
Since the interference pattern (Eq.~\eqref{eq:definition_of_interference_pattern}) satisfies $P\left(\boldsymbol{\phi}\right)\le 1+C_{l_{1}}$ for any value of $\boldsymbol{\phi}$, $C_{l_{1}}$ is bounded, $C_{l_{1}}(\rho)\geq\left|Q_n(\rho)\right|^{1/n}-1$, in terms of the moments $Q_n(\rho)$.  
It is thus not surprising, that $Q_{n}$ contains enough information to provide a valuable description of quantum coherence as well as practical criteria for identification of $k$-coherence.

\acknowledgements
We thank Alexander Stibor and Bj\"orn Witt for interesting and helpful conversations.
Financial support by the European Research Council under the project Odycquent is gratefully
acknowledged.

\appendix

\section{Derivation of threshold values\label{sec:Derivation-of-threshold-values}}
We start the derivation by inserting $\rho^{(k)}=\Ket{\Psi^{(k)}}\Bra{\Psi^{(k)}}$ with $\Ket{\Psi^{(k)}}$ given by
(\ref{kcohpure}) into the expression (\ref{eq:final-expression-for-p_n}):
\begin{eqnarray}
Q_{n} & = & \!\!\!\!\!\sum_{i_{1},i_{2},...,i_{2n}=1}^{d}\!\!\!\!\!\sqrt{\lambda_{i_{1}}\lambda_{i_{2}}...\lambda_{i_{2n}}}\prod_{i=1}^{d}R_{n_{i}}\cos X(\boldsymbol{\mu})\nonumber \\
 & \leq &\!\! \sum_{i_{1},i_{2},...,i_{2n}=1}^{d}\!\!\sqrt{\lambda_{i_{1}}\lambda_{i_{2}}...\lambda_{i_{2n}}}\prod_{j=1}^{d}R_{n_{j}}\nonumber \\
 &\equiv& g_{n}\left(\boldsymbol{\lambda}\right),\label{eq:p_n_res}
\end{eqnarray}
with
\begin{equation}
X(\boldsymbol{\mu})=\sum_{j=1}^{d}n_{j}\left(\mu_{j}-\varphi_{j}\right).
\end{equation}
An estimate $\cos\left(\cdot\right)\leq1$ applied in the second line
implies that the maximum of $Q_{n}$
with respect to $\boldsymbol{\mu}$ is achieved when the peak position
of the probability distribution coincides with the maximum of the
interference pattern. 

In the next step we employ the concept of Schur-concavity \cite{schur_concavity,schurscondition}.
For any two vectors $\boldsymbol{\lambda}$ and $\boldsymbol{\lambda}$'
such that $\boldsymbol{\lambda}'$ is majorized by $\boldsymbol{\lambda}$
(so that $\boldsymbol{\lambda}'\prec\boldsymbol{\lambda}$) and any Schur-concave
function $g\left(\boldsymbol{\lambda}\right)$ one gets $g\left(\boldsymbol{\lambda}'\right)\geq g\left(\boldsymbol{\lambda}\right)$.
In the case of pure, $k$-coherent states all vectors $\boldsymbol{\lambda}$
majorize the uniform vector
\begin{equation}
\boldsymbol{\lambda}_{k}=\left(\underbrace{\frac{1}{k},\frac{1}{k},...,\frac{1}{k}}_{k\textrm{-times}},\underbrace{0,...,0}_{\left(d-k\right)\textrm{-times}}\right).
\end{equation}
To finish the proof we thus only need to show that the function $g_{n}\left(\boldsymbol{\lambda}\right)$
defined in (\ref{eq:p_n_res}) is Schur-concave for $n=2$ and $n=3$. To this
end it is sufficient to show that $g_{n}\left(\boldsymbol{\lambda}\right)$
satisfies the condition
%\\
%\\
\begin{equation}
\textrm{S}_{ij}(g)=\left(\lambda_{i}-\lambda_{j}\right)\left(\frac{\partial g}{\partial\lambda_{i}}-\frac{\partial g}{\partial\lambda_{j}}\right)\leq0\,\,\,\,\forall\, i,j=1,2,...,d.
\end{equation}
%Since $g_{n}\left(\boldsymbol{\lambda}\right)$ is symmetric (under
%permutation of its arguments) we shall only prove this inequality for
%a single pair of indices, say $i=1$ and $j=2$.

To proceed further we need an explicit form of both functions; that is why we define
\begin{equation}
W_{A}^{B}=\prod_{l=1}^{A}\lambda_{i_{l}}\prod_{n=1}^{B}\sqrt{\lambda_{j_{n}}}\ ,
\end{equation}
and
\begin{equation}
G_{AB}=\sum_{\neq}W_{A}^{B}\ ,
\end{equation}
where $\sum_{\neq}$ denotes the sum over all pairwise different indices $i_1,\hdots,i_A,j_1,\hdots,j_B$ running from $1$ to $d$.
With this, one obtains
\begin{widetext}
\begin{equation}
g_{2}\left(\boldsymbol{\lambda}\right)=1+\left(1+R_{2}^{2}\right)G_{20}+2R_{1}^{2}G_{02}+2R_{1}^{2}\left(1+R_{2}\right)G_{12}+R_{1}^{4}G_{04},\label{eq:g2}
\end{equation}
and
\begin{eqnarray}
g_{3}\left(\boldsymbol{\lambda}\right) & = & 1+3R_{1}^{2}G_{02}+3\left(1+R_{2}^{2}\right)G_{20}+6R_{1}^{2}\left(1+R_{2}\right)G_{12}+3R_{1}^{4}G_{04}+2\left(1+3R_{2}^{2}\right)G_{30}+3R_{1}^{2}\left(3+4R_{2}+3R_{2}^{2}\right)G_{22}\nonumber \\
 & + &6R_{1}^{4}\left(1+R_{2}\right)G_{14}+ R_{1}^{6}G_{06}+6R_{1}\left(2R_{1}+R_{1}R_{2}+R_{2}R_{3}\right)\sum_{\neq}\sqrt{\lambda_{i}}^{3}\lambda_{j}\sqrt{\lambda_{m}}\label{eq:g3}\\
 & + & 2R_{1}^{3}\left(3R_{1}+R_{3}\right)\sum_{\neq}\sqrt{\lambda_{i}}^{3}\sqrt{\lambda_{j}\lambda_{m}\lambda_{l}}+\left(3R_{1}^{2}+R_{3}^{2}\right)\sum_{\neq}\sqrt{\lambda_{i}\lambda_{j}}^{3}\ .\nonumber 
\end{eqnarray}
%\end{widetext}
%where the symbol $\sum_{\neq}$ represents the sum from $1$ to $d$ with all indices being pairwise different.

Since all parameters $R_{n}$ are non-negative we can treat each term
in (\ref{eq:g2}-\ref{eq:g3}) separately and find (for $i\neq j$)
%\begin{widetext}
\begin{eqnarray}
\frac{\partial G_{AB}}{\partial\lambda_{i}}&=&A\sum_{\neq,\setminus\set{i}}W_{A-1}^{B}+\frac{B}{2\sqrt{\lambda_{i}}}\sum_{\neq,\setminus\set{i}}W_{A}^{B-1}\\
&=&
A\sum_{\neq,\setminus\set{i,j}}\left(W_{A-1}^{B} +(A-1)\lambda_{j} W_{A-2}^{B} + B\sqrt{\lambda_{j}} W_{A-1}^{B-1}\right)\nonumber\\ 
&+& \frac{B}{2\sqrt{\lambda_{i}\lambda_{j}}} \sum_{\neq,\setminus\set{i,j}} \left(\sqrt{\lambda_{j}}W_{A}^{B-1}+A\lambda_{j}^{3/2}W_{A-1}^{B-1}+(B-1)\lambda_{j} W_{A}^{B-2}\right)\ ,
\label{eq:app1}
\end{eqnarray}
\end{widetext}
where $\sum_{\neq,\setminus Z}$ denotes the sum $\sum_{\neq}$ with the additional exclusion of the values $Z$.
From Eq.~\eqref{eq:app1}
one may see that $\textrm{S}_{ij}(G_{AB})$  (with $i\neq j$) contains only terms proportional to $-\left(\lambda_{i}-\lambda_{j}\right)^{2}$, $\left(\lambda_{i}-\lambda_{j}\right)(\sqrt{\lambda_j}-\sqrt{\lambda_i})$ and  $\left(\lambda_{i}-\lambda_{j}\right)(\lambda_j^{3/2}-\lambda_i^{3/2})$.
Since all these terms are separately non-positive, 
$g_{2}\left(\boldsymbol{\lambda}\right)$ is a sum of Schur-concave functions $G_{AB}$, and thus Schur concave.

The function $g_{3}\left(\boldsymbol{\lambda}\right)$
involves additional terms which are not of the form of $G_{AB}$ and
are not Schur-concave.
It is, however, possible to show that the terms
\begin{equation}
2R_{1}^{3}\left(3R_{1}+R_{3}\right)\sum_{\neq}\sqrt{\lambda_{i}}^{3}\sqrt{\lambda_{j}\lambda_{m}\lambda_{l}}+6R_{1}^{4}\left(1+R_{2}\right)G_{14},
\end{equation}
\begin{eqnarray*}
\textrm{and}\quad 6R_{1}\left(2R_{1}+R_{1}R_{2}+R_{2}R_{3}\right)\sum_{\neq}\sqrt{\lambda_{i}}^{3}\lambda_{j}\sqrt{\lambda_{m}}+\\ \left(3R_{1}^{2}+R_{3}^{2}\right)\sum_{\neq}\sqrt{\lambda_{i}\lambda_{j}}^{3}
+\,3R_{1}^{2}\left(3+4R_{2}+3R_{2}^{2}\right)G_{22},
\end{eqnarray*}
are Schur-concave for $R_{1}\geq R_{3}$.
For the wrapped normal distribution we get $R_{1}=e^{-\sigma^{2}/2}\geq e^{-9\sigma^{2}/2}=R_{3}$,
so that $g_{3}\left(\boldsymbol{\lambda}\right)$ is Schur-concave
as well.

\section{Derivation of the global maximum \label{sec:Derivation-of-maximum}}

Since the generalized moments (\ref{eq:final-expression-for-p_n})
are real and non-negative we have the estimate
\begin{equation}\label{B1}
Q_{2}=\left|Q_{2}\right|\leq\sum_{i_{1},i_{2},i_{3},i_{4}=1}^{d}\left|\rho_{i_{1}i_{3}}\right|\left|\rho_{i_{2}i_{4}}\right|\prod_{j=1}^{d}R_{n_{j}}.
\end{equation}
Since the right hand side does not depend on $\boldsymbol{\mu}$,
the same upper bound applies to the maximum of $\left|Q_{2}\right|$ (maximized with respect to the center of the distribution $F(\boldsymbol{\boldsymbol{\phi}})$).

The right hand side of (\ref{B1}) explicitly reads
\begin{eqnarray}
1+\left(1+R_{2}^{2}\right)\sum_{\neq}\left|\rho_{ij}\right|^{2}+2R_{1}^{2}\sum_{\neq}\left|\rho_{ij}\right|+R_{1}^{4}\sum_{\neq}\left|\rho_{ij}\right|\left|\rho_{kl}\right|\nonumber \\
 + \,2R_{1}^{2}\left(1+R_{2}\right)\sum_{\neq}\left|\rho_{im}\right|\left|\rho_{ji}\right|,\qquad
\end{eqnarray}
 and saturates if
$\mathbb{R}\ni\rho_{ij}\geq0$ for all $i\neq j$, so that the density
matrix has only real and non-negative entries. 

Using the inequality of arithmetic and geometric mean we get the
estimates
\begin{equation}
\sum_{\neq}\left|\rho_{im}\right|\left|\rho_{ji}\right|\leq\left(d-2\right)\sum_{\neq}\left|\rho_{ij}\right|^{2},\label{13}
\end{equation}
\begin{equation}
\sum_{\neq}\left|\rho_{ij}\right|\left|\rho_{ml}\right|\leq\left(d-3\right)\left(d-2\right)\sum_{\neq}\left|\rho_{ij}\right|^{2}.\label{13-1}
\end{equation}
Since we assume that the purity $\mathcal{P}=\textrm{Tr}\rho^{2}$
is fixed we get
\begin{equation}
\sum_{\neq}\left|\rho_{ij}\right|^{2}=\mathcal{P}-\sum_{i=1}^{d}\rho_{ii}^{2},\label{Identity}
\end{equation}
what also implies that
\begin{equation}
\sum_{\neq}\left|\rho_{ij}\right|\leq\sqrt{d\left(d-1\right)\left(\mathcal{P}-\sum_{i=1}^{d}\rho_{ii}^{2}\right)}.\label{16}
\end{equation}

The maximum of (\ref{Identity}), as well as maxima of (\ref{13},
\ref{13-1}) and (\ref{16}) are provided by the uniform distribution
\begin{equation}
\forall_{i}\,\rho_{ii}=\frac{1}{d}.
\end{equation}
This observation immediately leads to a state independent maximum
of $Q_2$. The above maximum may
be saturated only when the inequalities used in (\ref{13}, \ref{13-1})
and (\ref{16}) saturate too, i.e. when 
\begin{equation}
\left|\rho_{ij}\right|=\frac{1}{d}\sqrt{\frac{\mathcal{P}d-1}{\left(d-1\right)}}\ ,\ \forall_{i\neq j}\ .
\end{equation}
The last conclusion proves Eq. (\ref{state1}), showing that the maximum
is always global. 

The next step is to determine when the state (\ref{state1}) does
not happen to be $k+1$-coherent. For example, if $\mathcal{P}=1$,
the global maximum is attained by a $d$-coherent pure
state. We start with the following observation: to realize any $k$-coherent state
it is sufficient to consider the form:
\begin{equation}
\rho^{(k)}=\sum_{m=1}^{D_{k}}\sum_{i,j\in I_{m}\left(k\right)}\Xi_{ij}^{\left(m\right)}\left|i\right\rangle \left\langle j\right|,\qquad D_{k}=\left(\begin{array}{c}
d\\
k
\end{array}\right),\label{dec2}
\end{equation}
with $I_{m}\left(k\right)$ being for each $m=1,\ldots,D_{k}$ a unique
set of $k$ different indices taken from $\left\{ 1,\ldots,d\right\} $.
By $\Xi_{ij}^{\left(m\right)}$ we denote arbitrary (not normalized)
$k\times k$ positive semi-definite matrices.

We shall now construct $\rho^{(k)}$ with all diagonal elements equal to
$\kappa\equiv1/d$ and all off-diagonal elements equal to $0\leq\beta\leq1/d$.
The maximal value of $\beta$ is provided by the case when for every
$m$, all diagonal elements of $\Xi_{ij}^{\left(m\right)}$ are equal
to some $h>0$ and all corresponding off-diagonal elements are given
by some $b\leq h$. From combinatorial considerations we find
\begin{equation}
\frac{1}{d}\equiv\kappa=\left(\begin{array}{c}
d-1\\
k-1
\end{array}\right)h,\qquad\beta=\left(\begin{array}{c}
d-2\\
k-2
\end{array}\right)b.
\end{equation}
We thus obtain
\begin{eqnarray}
\textrm{Tr}\left(\rho^{(k)}\right)^{2} & = & \frac{1}{d}+d\left(d-1\right)\beta^{2}\nonumber \\
&=&\frac{1}{d}+d\left(d-1\right)\left[\left(\begin{array}{c}
d-2\\
k-2
\end{array}\right)\right]^{2}b^{2}\nonumber \\
 & \leq & \frac{1}{d}+d\left(d-1\right)\left[\left(\begin{array}{c}
d-2\\
k-2
\end{array}\right)\right]^{2}h^{2}\nonumber \\&=&\mathcal{P}_{k}.
\end{eqnarray}
In that way we have recovered the range given in Eq.~\eqref{range1}.

%\section*{References}{\bibliographystyle{unsrt}
%\addcontentsline{toc}{section}{\refname}\bibliography{references}
%}
\end{document}